\documentclass{emulateapj}

\begin{document}

\title{The Transformation of Cluster Galaxies at Intermediate Redshift\thanks{Based on observations carried out at the ESO Very
Large Telescope (VLT), Chile, as part of the ESO Large Program
LP-166.A-0701 and proposals 69.A-0683 and 72.A-0759.}}

\author{
N.~L. Homeier\altaffilmark{1} 
R. Demarco\altaffilmark{1},
P. Rosati\altaffilmark{12},
M. Postman\altaffilmark{1},
J.~P. Blakeslee\altaffilmark{1}, 
R.~J. Bouwens\altaffilmark{3},
L.~D. Bradley\altaffilmark{1},
H.~C. Ford\altaffilmark{1},
T. Goto\altaffilmark{1},
C. Gronwall\altaffilmark{9},
B. Holden\altaffilmark{3},
G.~D. Illingworth\altaffilmark{3},
M.~J. Jee\altaffilmark{1},
A.~R. Martel\altaffilmark{1},
S. Mei\altaffilmark{1},
F. Menanteau\altaffilmark{1},
A. Zirm\altaffilmark{8},
M. Clampin\altaffilmark{7},
G.~F. Hartig\altaffilmark{2},
D.~R. Ardila\altaffilmark{1},
F. Bartko\altaffilmark{4}, 
N. Ben\'{\i}tez\altaffilmark{1},
T.~J. Broadhurst\altaffilmark{5},
R.~A. Brown\altaffilmark{2},
C.~J. Burrows\altaffilmark{2},
E.~S. Cheng\altaffilmark{6},
N.J.G. Cross\altaffilmark{1},
P.D. Feldman\altaffilmark{1},
M. Franx\altaffilmark{8},
D.~A. Golimowski\altaffilmark{1},
L. Infante\altaffilmark{10}
R.~A. Kimble\altaffilmark{7},
J.~E. Krist\altaffilmark{2},
M.~P. Lesser\altaffilmark{11},
G.~R. Meurer\altaffilmark{1},
G.~K. Miley\altaffilmark{8},
V. Motta\altaffilmark{10},
M. Sirianni\altaffilmark{2}, 
W.~B. Sparks\altaffilmark{2}, 
H.~D. Tran\altaffilmark{13}, 
Z.~I. Tsvetanov\altaffilmark{1},   
R.~L. White\altaffilmark{2},
\& W. Zheng\altaffilmark{1}}


\altaffiltext{1}{Department of Physics and Astronomy, Johns Hopkins
University, 3400 North Charles Street, Baltimore, MD 21218.}

\altaffiltext{2}{STScI, 3700 San Martin Drive, Baltimore, MD 21218.}

\altaffiltext{3}{UCO/Lick Observatory, University of California, Santa
Cruz, CA 95064.}


\altaffiltext{4}{Bartko Science \& Technology, 14520 Akron Street, 
Brighton, CO 80602.}	


\altaffiltext{5}{Racah Institute of Physics, The Hebrew University,
Jerusalem, Israel 91904.}


\altaffiltext{6}{Conceptual Analytics, LLC, 8209 Woburn Abbey Road, Glenn Dale, MD 20769}


\altaffiltext{7}{NASA Goddard Space Flight Center, Code 681, Greenbelt, MD 20771.}


\altaffiltext{8}{Leiden Observatory, Postbus 9513, 2300 RA Leiden,
Netherlands.}


\altaffiltext{9}{Department of Astronomy and Astrophysics, The
Pennsylvania State University, 525 Davey Lab, University Park, PA
16802.}


\altaffiltext{10}{Departmento de Astronom\'{\i}a y Astrof\'{\i}sica,
Pontificia Universidad Cat\'{o}lica de Chile, Casilla 306, Santiago
22, Chile.}


\altaffiltext{11}{Steward Observatory, University of Arizona, Tucson,
AZ 85721.}


\altaffiltext{12}{European Southern Observatory,
Karl-Schwarzschild-Strasse 2, D-85748 Garching, Germany.}


\altaffiltext{13}{W. M. Keck Observatory, 65-1120 Mamalahoa Hwy., 
Kamuela, HI 96743}

\altaffiltext{8}{Leiden Observatory, Postbus 9513, 2300 RA Leiden,
Netherlands.}

\begin{abstract}
We combine imaging data from the Advanced Camera for Surveys (ACS) with 
VLT/FORS optical spectroscopy to study the properties of star-forming galaxies
in the $z=0.837$ cluster CL0152-1357. We have morphological information for 24 
star-forming cluster galaxies, which range in morphology 
from late-type and irregular
to compact early-type galaxies. We find that while most star-forming galaxies
have $r_{625}-i_{775}$ colors bluer than 1.0, eight are in the red cluster sequence. 
Among the star-forming cluster population we find five compact early-type 
galaxies which have properties consistent with their identification as
progenitors of dwarf elliptical galaxies.
The spatial distribution of the star-forming cluster members is nonuniform. 
We find none within $R\sim 500$~Mpc of the cluster center, which is
highly suggestive of an intracluster medium interaction. We derive star formation
rates from [OII]~$\lambda\lambda 3727$ line fluxes, and use these to compare the
global star formation rate of CL0152-1357 to other
clusters at low and intermediate redshifts. 
We find a tentative correlation between integrated star
formation rates and $T_{X}$, in the sense that hotter clusters have lower
integrated star formation rates. Additional data from clusters 
with low X-ray temperatures is needed to confirm this trend.
We do not find a significant correlation with redshift,
suggesting that evolution is either weak or absent between $z=0.2-0.8$.
\end{abstract}


\section{Introduction}

Galaxy morphology is influenced by environment. 
This is illustrated at the most basic level by the 
difference in the morphological mix of galaxies
in high density galaxy clusters and low density groups, 
or the field. In the local universe, the highest 
density environments of massive cluster cores 
are composed almost exclusively of early-type galaxies,
i.e. ellipticals and lenticulars (S0s), while in the field,
gas-rich disks and irregular galaxies dominate. 
In the lower density field and in low-mass groups, interactions and 
mergers are linked to enhanced star formation rates
and starbursts 
\citep{S87,SM96,Jetal99,CBG00,BGK00,Petal01,Hetal02,SS03}. 
It might be expected that ram pressure from a dense intracluster medium 
(ICM) compresses molecular clouds and triggers a starburst \citep{BekkiC03}.
Therefore, it is not intuitively
clear if gas-rich galaxies infalling into high density clusters 
would have 
increased or decreased star formation rates relative to the field. 
In intermediate redshift cluster samples, the star formation
rates of cluster galaxies are {\it suppressed} relative to the field,
even when the different morphological mix is accounted for
\citep{Baloghetal97,Betal98,Hashetal98}. While the star formation rates 
in clusters increase as a function
of redshift, they are still lower than the star formation rates
in the field population
at that redshift. The mechanism by which star formation 
in cluster galaxies is
suppressed is still unclear, but there is substantial 
evidence that suppression, not enhancement, is the dominant outcome
\citep{Cetal01,Ellingsonetal01,PLO01,Lewisetal02,Metal02}.


Two of the most widely discussed 
mechanisms for star formation suppression
are ram pressure stripping \citep{GG72}, and a more gentle form of galaxy-ICM
interaction termed 'starvation' \citep{LTC80}. Both of these mechanisms
may operate with varying degrees of importance, but stripping is 
expected to quench star formation rapidly ($\sim 50$~Myr; e.g. Quilis, Moore, \&
Bower 2000), 
whereas starvation leads to a decline in the star formation rate over a 
longer timescale ($> 1$~Gyr; e.g. Bekki, Couch, \& Shioya 2002).
The other mechanism of note is galaxy harassment, where
galaxy-galaxy interactions combined with the tidal field of the
cluster are expected to lead to strong mass loss and a drastic change 
in morphology \citep{MLK98}. 

In the Virgo cluster, ram pressure stripping is the leading 
explanation for the observation that gas-deficient dwarf
galaxies are found in regions of higher X-ray surface brightness
\citep{Letal00}.
Also in Virgo, there is direct evidence for ram pressure
stripping of spiral galaxies in the form of HI and $H\alpha$
``bow-shock'' morphology \citep{Vetal04,Ketal04}.
Further evidence for the importance of galaxy-ICM interactions
comes from a study by Smith (2003), who
found that the hot X-ray halos of cluster early-type galaxies are
missing, showing that
stripping of gas halos occurs in dense environments.
Also, the existence of cluster passive spirals, 
galaxies with spiral structure but no
detected star formation, can also be interpreted as evidence
for ram pressure strippping \citep{Gotoetal03,YG04,Kodamaetal04}.
However, the 
importance of ram pressure stripping and starvation 
on the morphology-density
relation as a whole has yet to be determined.
Due to degenerate outcomes in color-morphology space
of vastly different formation
scenarios, we must observe at higher redshifts to 
probe the dominant processes in 
producing the morphology-environment correlations seen in the 
present day universe.

CL0152-1357 ($z=0.837$) was observed as part of an ongoing project with
the Advanced Camera for Surveys (ACS) to obtain
imaging of intermediate redshift cluster 
galaxies, with the goal of tracing the 
evolution of cluster galaxy properties.
In this paper we combine morphological
information from ACS imaging with ground-based spectroscopy to
characterize the population of star-forming cluster galaxies,
and to provide constraints on the mechanisms which influence the
evolution of the star-forming population in clusters.



We begin our investigation of how the properties of star-forming
galaxies depend on environment in \S~3 by establishing the morphologies, 
colors, and locations of our sample galaxies. In \S~4 we calculate
star formation rates for individual galaxies, and use these to
derive an integrated star formation rate for the cluster. We compare 
this to other clusters at low to intermediate redshift in \S~5,
and explore relationships with redshift, $L_{X}$, and $T_{X}$.
In \S~6 we discuss the possibility that ram pressure stripping is 
responsible for extinguishing star formation in CL0152-1357, and in
\S~7 we discuss the likely fate of the star-forming galaxies.
In \S~8, we present our conclusions.

We use $\Omega_{m}=0.3, \Omega_{\Lambda}=0.7$, and 
$H_{0}=70$~km~s$^{-1}$.

\section{Observations and Reductions}

CL$0152-1357$ was discovered in three cluster surveys: the 
Wide Angle {\it ROSAT} Pointed
Survey (WARPS; Ebeling et~al. 2000), the {\it ROSAT} Deep Cluster Survey 
(RDCS, Rosati et~al. 1998) and the Serendipitous High-redshift Archival 
{\it ROSAT} Cluster Survey (SHARC; Romer et~al.~2000). 
It is not dynamically relaxed, 
and contains two massive clusters in the process of merging,
both at redshifts of $z\sim0.837$ \citep{Demarco03,Demarcoetal04}. 
The mass within 1~Mpc is $4.9\pm0.4\times 10^{14}$M$_{\odot}$ from
a weak-lensing analysis \citep{Jeeetal04}.
There also appears to be
structure in addition to the two massive components. 
An example of this is a group of red cluster sequence galaxies to the east of the northern
component; this structure is also detected in the weak-lensing mass map as
well as diffuse X-ray emission.

CL0152-1357 was observed in the F625W, F775W, and F850LP bandpasses
(hereafter $r_{625}$, $i_{775}$, and $z_{850}$) with the ACS \citep{Fordetal02}
Wide Field Channel
as part of the guaranteed time observation program (proposal 9290).
The observations were taken in a $2\times2$ mosaic pattern, with 2 
orbits of integration in the $r_{625}$, $i_{775}$, and $z_{850}$ filters 
at each of the four pointings. Due to the $1\arcmin$ of
overlap between the pointings, the core of the cluster was imaged for 
a total of 8 orbits in each filter.
The data were processed with the $Apsis$ pipeline \citep{Betal03a,Betal03b}. 
Our photometry is calibrated to the AB
magnitude system using zeropoints of $25.90\pm0.04$ ($r_{625}$), 
$25.66\pm0.03$ ($i_{775}$), and 
$24.86\pm0.02$ ($z_{850}$) (Sirianni et~al., in preparation). 
Object detection and photometry was performed by SExtractor \citep{BA96}
incorporated within the $Apsis$ pipeline. A more detailed description can be
found in \citet{Benitez04}. We use isophotal magnitudes
(MAGISO) for colors and total magnitudes (MAGAUTO) from {\tt Sextractor} 
\citep{BA96}. 

\begin{figure}
\begin{center}
\includegraphics[width=16cm]{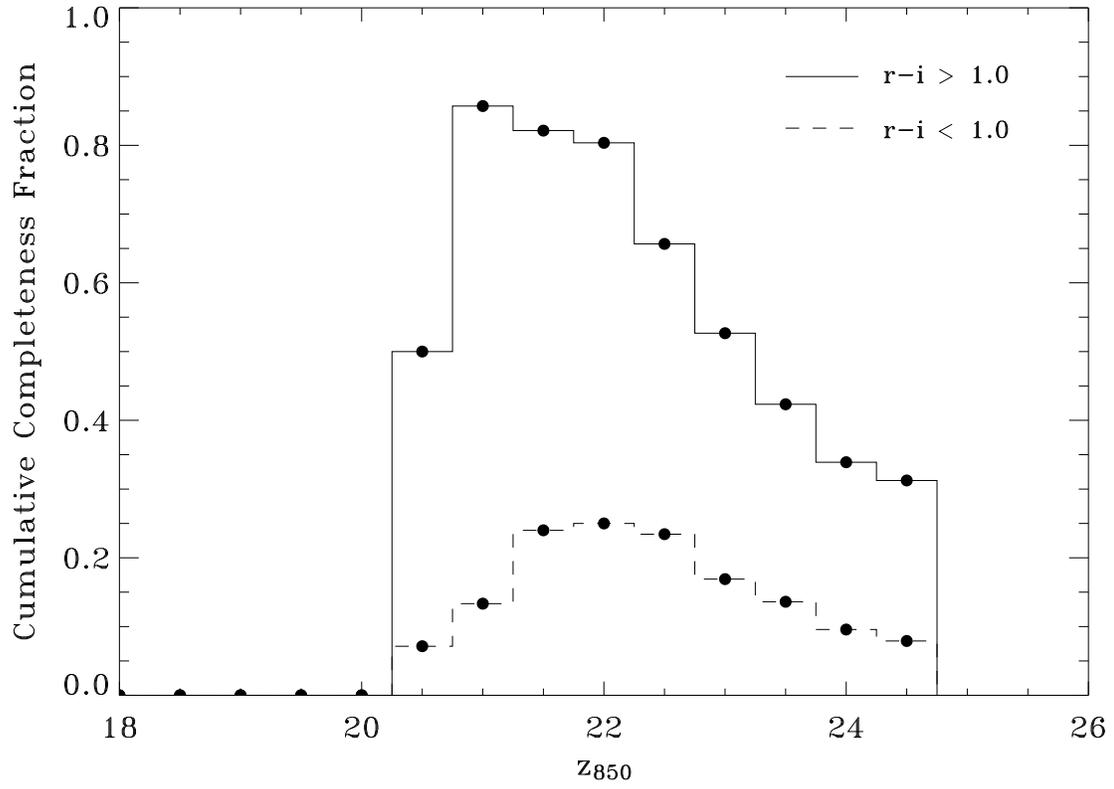}
\caption{Cumulative redshift completeness for blue ($r-i < 1.0$)
and red ($r-i > 1.0$) galaxies.
\label{selfunc}}
\end{center}
\end{figure}

The spectroscopic data were taken with the FORS1 and FORS2 instruments
on the VLT, and are fully described in Demarco et~al. The pixel scale 
ranges
from $0.20-0.25\arcsec$ pixel$^{-1}$, the spectral resolution
from $2.6-6.9$~\AA~pixel$^{-1}$, the slit widths from 
$1-1.4\arcsec$, and the seeing was $<0.8\arcsec$ (Demarco et al. 2005). 
The wavelength coverage was $4000-10000$~\AA or $6000-10000$~\AA.
S/N ranged from less than 1 in cases where no continuum was detected, to
$\sim 50$ at $4000$~\AA~rest-frame.
Redshift completeness fractions were determined by counting the 
number of
galaxies in the field of view within the specified magnitude (color, 
morphology) 
limits compared to the fraction of those with redshifts. The sample of 
100 field galaxies used 
in \S~3.1 is from the spectroscopic sample of confirmed
non-cluster members with redshifts and ACS coverage. This field sample
covers the redshift range $0.2 < z < 1.3$. Galaxies for
spectroscopic follow-up were selected as having photometric
redshifts $0.7 < z_{phot} < 0.9$. The redshift 
selection was not uniform with color, but was biased towards red galaxies.
Therefore, the completeness as a function of color, shown in 
Figure~\ref{selfunc}, varies by a 
factor of 2 for blue and red galaxies brighter than $i=23$.

\section{Results}

\begin{figure*}
\begin{center}
\includegraphics[width=10cm]{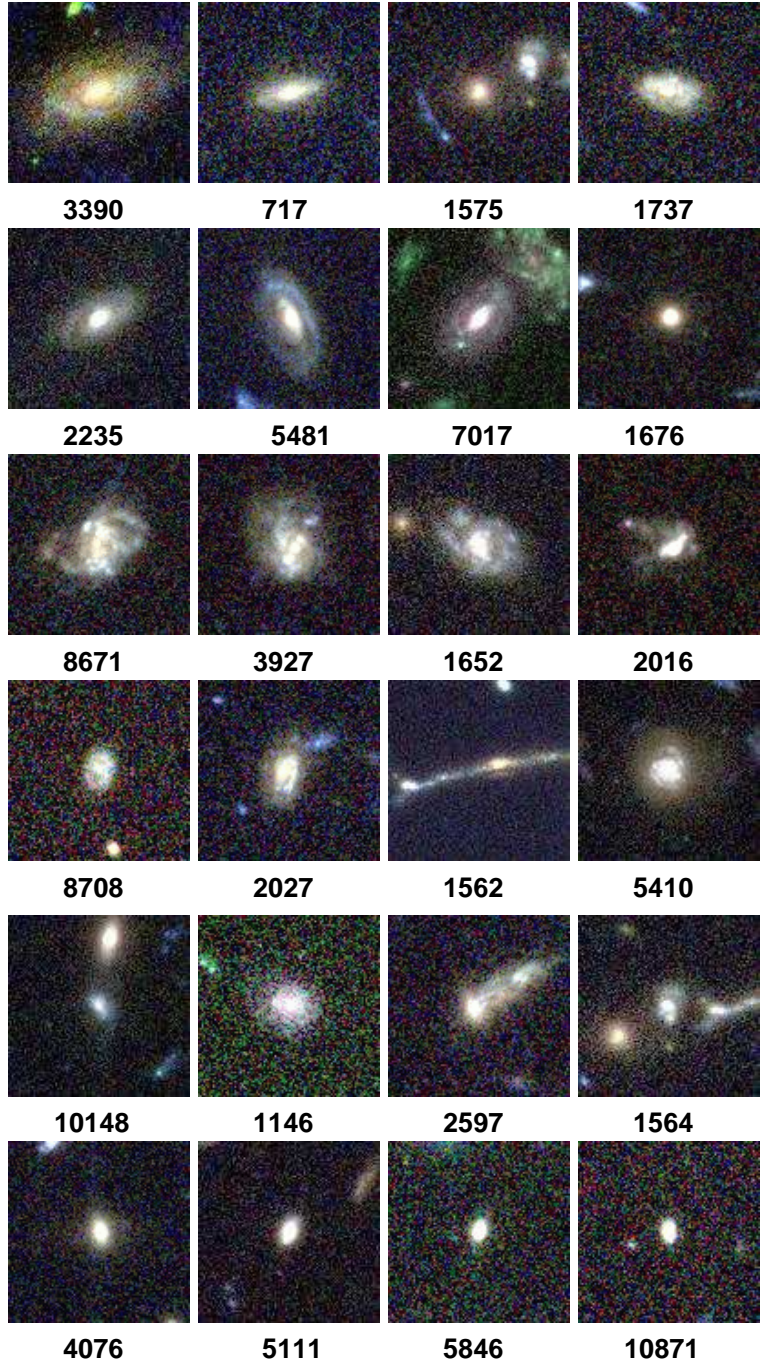}
\caption{ACS $r_{625}, i_{775}, z_{850}$ color cutouts of the spectroscopically
confirmed star-forming cluster members. 
Top two rows: red cluster sequence members, middle three rows: spirals 
and irregulars, bottom row: compact early-type galaxies. Galaxy 1676 is
a compact early-type, but also in the red sequence.
ACS IDs corresponding to Table~\ref{sftab} are given below
each cutout. 
Image sizes are $5\arcsec\times 5\arcsec$.
\label{sfrgalaxies}}
\end{center}
\end{figure*}

The spectroscopic sample contains 102 cluster members, 33 
of which have emission lines indicative of
star formation, and 2 are AGN. Emission lines were
identified by eye on the 2D spectral images. 
The weakest [OII] line identified
is 12~\AA~EW. Table~\ref{sftab} lists
information for the star-forming galaxies (excluding AGN).
We have morphological information for 24
of the star-forming galaxies that overlap with our ACS field. 
Figure~\ref{sfrgalaxies} shows $r_{625}, i_{775}, z_{850}$ color
cutouts of the star-forming cluster members.

\subsection{Morphology}

Morphologies were determined by visual classifications
on the T-type system \citep{deVacetal91}. All galaxies in
the field with $i_{775} \le 24$ magnitude were classified
by MP (Postman et al., in preparation). 
20\% of the galaxies were also classified by MF, NC, and BH 
to estimate the classification errors. Unanimous or
majority agreement was achieved for 75\% of objects with 
$i_{775}\le 23.5$. There was no significant offset between
the mean classification from the independent classifiers.

Visual morphologies of the cluster 
members with evidence for ongoing star formation, indicated by the
detection of [OII] emission, fall into three categories: 
red cluster sequence
{\it spiral} galaxies (Sa and earlier), compact ($r_{e} < 3$~kpc) 
early-type galaxies ($T \le 0$), 
and irregular/late-type spiral galaxies ( $T \ge 3$) that 
are typically expected to have star formation. 
Only one of the galaxies (2016) is highly disturbed, 
indicative of a major merger.

\begin{figure*}
\begin{center}
  \includegraphics[width=16cm,scale=0.5]{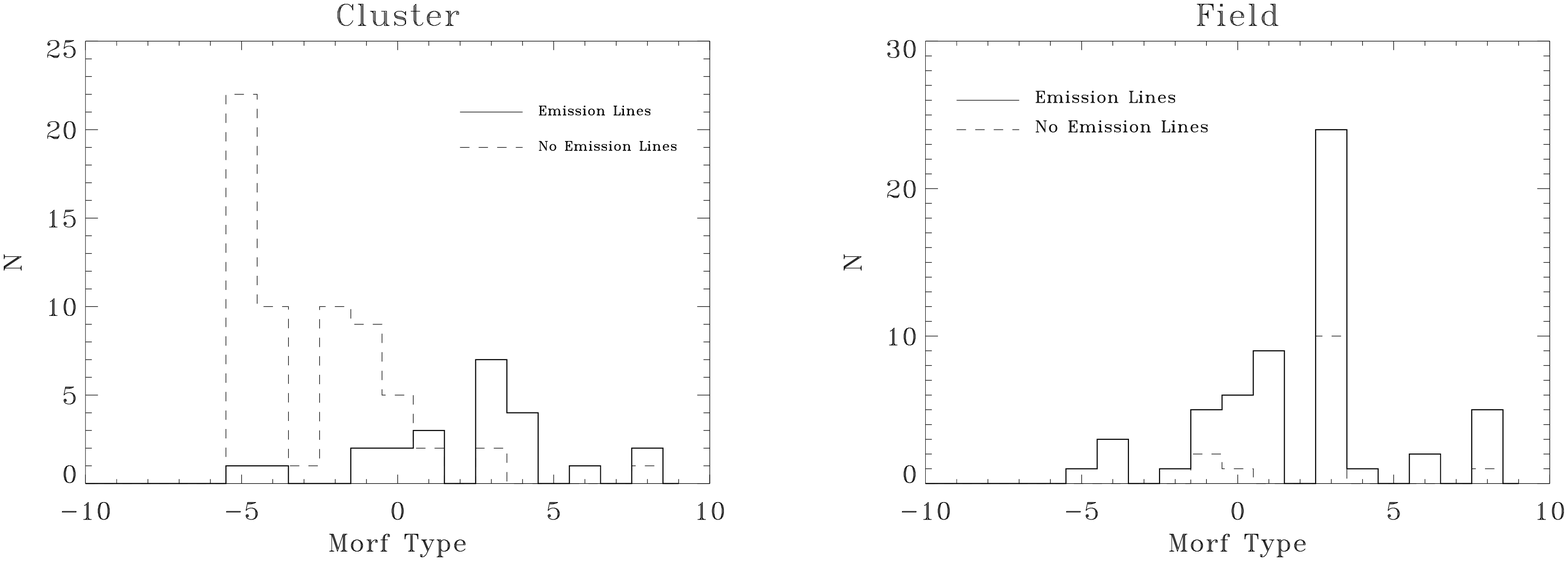}
  \caption{The distribution of visually classified morphologies for 
galaxies with and without emission lines. There is a clear trend 
for cluster galaxies without [OII] to have early-type morphologies, 
while the star-forming cluster galaxies range from early to late-type.
This trend is absent in the field sample. 
    \label{morfhist}}
\end{center}
\end{figure*}

In Figure~\ref{morfhist} we present histograms
of the visually classified morphologies for galaxies with 
and without emission
lines. For the cluster galaxies, we
see a clear correlation of morphology with star formation, which
is absent in the field sample. Also, the distribution of morphological
types is different between cluster and field, with a significant number 
of disky galaxies in the field sample, and early-type morphologies in
the cluster sample.

Five of the star-forming cluster members have early-type
morphologies and compact appearances. We fit 2D 
light profiles to the PSF-convolved
$z_{850}$ images using GALFIT \citep{Petal02}. The results 
are shown in Table~\ref{tab:compacts}. We fit a single
Sersic profile where the surface brightness varies as
$r^{1/n}$ \citep{S68,CB99}.
A profile with $n=4$ is a de Vaucouleurs profile, and one
with $n=1$ describes an exponential disk.
We find that these compact early-type cluster members have 
effective radii ranging from $\sim 1-2.7$~kpc. This is lower
than the mean of 3.4~kpc from 53 early-type (T $\le 0$)
cluster member galaxies. Thus, we are able to confirm their 
visually compact appearance.
The low measured $n$ values for these compact
early-type galaxies is consistent the trend of magnitude
and surface brightness profile found for E and dE galaxies 
in the nearby Universe \citep{GG03}.

None of the early-type galaxies with [OII] is a luminous
point source in the 37ks {\it Chandra} image.
These galaxies have narrow [OII] $\lambda 3727$
emission lines, and none has [NeIII] $\lambda 3869$; broad lines
and [NeIII] $\lambda 3869$ 
are typical of active nuclei. For galaxy 1006=10871, we also
detect [OIII] $\lambda 5007$ and H$\beta$, and we can apply 
a diagnostic criteria from \citet{Rolaetal97}. We find
log([OIII]$\lambda 5007$/H$\beta$)=0.3 and 
log([OII]$\lambda 3727$/H$\beta$)=0.6, locating it 
within the HII galaxy regime, but in a region also
populated by LINERs. Thus we cannot rule out
an active nucleus origin for the [OII] emission. This applies
to all galaxies in our sample, however, these compact 
early-type galaxies lack properties typical of star-forming galaxies, 
such as a patchy appearance (indicating the presence of gas and dust).

\subsection{Colors of Star-forming Galaxies}

In Figure~\ref{cmd} we present a histogram of the $r_{625}-i_{775}$
values for cluster member galaxies with and without detected star formation.
There is a strong segregation in the $r_{625}-i_{775}$ colors of 
star-forming and non-star-forming galaxies. While the star-forming
galaxies range in color from blue to red, galaxies without detected
star formation are almost exclusively red. This leads to the
conclusion that with reliable photometric
redshifts, a simple color selection to separate the star-forming and
passive cluster populations is fairly robust (e.g. Gray et~al. 2004). 
However, such a selection is not well suited to study the overall
properties of the star-forming cluster members; there is a small
but significant fraction of red star-forming galaxies.

\subsubsection{Fraction of Star-forming Red Cluster Sequence Galaxies}

The red cluster sequence (RCS) is the prominent red ridge in
color magnitude diagrams of large galaxy samples. In low redshift clusters,
it is composed exclusively of ellipticals and S0s. 
However in this galaxy sample we have RCS members with
current star formation as indicated by the 
presence of [OII] emission, also noted in \citet{Demarcoetal04}. 
To calculate the fraction 
of RCS galaxies with ongoing star formation, we need the 
completeness of our spectroscopic survey as a function of 
magnitude, and the magnitude distribution of the red cluster sequence (RCS)
galaxies. The redshift completeness as a function of color
is shown in Figure~\ref{selfunc}. Galaxy selection was biased towards 
red galaxies. After correcting for color, there was also a $\sim5$\% 
bias towards early-type morphologies, but we will neglect this correction
as it is much smaller than the color bias.

To define the red cluster sequence we take all non-stellar objects with
$0.8 < r_{625}-i_{775} < 1.4$ and $20 < i_{775} < 24$. We iteratively
fit a linear relation to this sample with $2.2\sigma$ clipping. 
The color-magnitude relation 
defining the red cluster sequence for CL0152-1357 is thus 

\begin{eqnarray*}
 (r-i) = (-0.037\pm 0.006)\times i + 1.900\pm 0.132
\end{eqnarray*}

The standard deviation about this relation is 0.07 magnitudes, and 
we define an object as an RCS member if its deviation is less than 
$2\sigma$, or $0.15$ magnitudes, 
from this relation. A more detailed
explanation of the red cluster sequence will be presented in 
Blakeslee et~al. (in preparation).


The deviation in magnitudes from the RCS, $\Delta$~RCS, is given in column 
8 of Table~\ref{sfrs}. There are 8 galaxies
with colors placing them in the RCS.
Of the RCS galaxies with spectra, how many show evidence for
star formation? There are spectra for 73 RCS cluster member galaxies. 
Of these 73, 44 are brighter than $i_{775}=22.5$, and 6 of the 44
have evidence for star formation ($\sim14$\%). If
we choose a slightly fainter magnitude limit of $i_{775}=23.0$,
then the number of RCS cluster member galaxies with spectra is 65, 
and 8 have detected [OII] ($\sim 12$\%).

After correcting for color bias in the redshift 
incompleteness fractions, there is a slight bias 
against galaxies with late-type morphologies 
(T$\ge 0$).
We conclude that the fraction of RCS galaxies with ongoing star formation
in CL$0152-1357$ is $>0.15$. We note that
this is an underestimate, because we have access only to
[OII] in emission as our star formation indicator.

\subsection{Segregation of Star-forming Galaxies}

In Figure~\ref{pos} we show the {\it Chandra} contour map
with the positions of the cluster members indicated as 
solid dots, and the edges of the ACS mosaic as dotted lines.
The star-forming galaxies are marked according to their 
morphology: triangles, squares, and diamonds represent
compact galaxies, spirals/irregulars, and RCS spirals.  
A similar figure without the morphological
imformation can be found in \citet{Demarcoetal04}.
We find no star-forming galaxies within the main X-ray
peaks, that is, within a contour value of 0.3 counts s$^{-1}$.
Considering the area between the two main X-ray peaks as the
center, there appears to be a sharp transition around $R\sim500$~kpc
($R\sim 300-400$~kpc from the center of the southern cluster), 
within which we find almost no star-forming galaxies. We note
that if the distribution of cluster galaxies is spherically symmetric,
a few might be expected simply from projection effects.
However, there are also few spectroscopically confirmed cluster
members within $\sim 500$~kpc to the northwest and southeast of the 
northern cluster. Therefore, the radius at which this transition 
occurs is not well constrained by our observations.

\section{Star Formation Rates}

Star formation rates are derived from [\ion{O}{2}] 
emission line luminosities. There are a few prescriptions 
for converting
[OII] line fluxes to star formation rates (Gallagher, 
Hunter, \& Bushouse 1989, Kennicut 1992,
Kewley, Geller, \& Jansen 2004). We will use the
prescription of Kewley et~al. (2004),
\begin{eqnarray*}
SFR_{[OII]} (M_{\odot}~yr^{-1}) = \\(6.58\pm1.65)\times 10^{-42} L_{[OII]} (ergs~s^{-1})
\end{eqnarray*}

This relation was derived using the mean reddening corrected
[OII]/H$\alpha$ ratio ($1.2\pm 0.3$) from the Nearby Field 
Galaxies Survey \citep{Jansenetal00}. Systematic errors
in [OII] derived SFRs arise from metallicity and extinction
effects. For field galaxies in the nearby Universe, both of these quantities have
been shown to correlate with $B$-band luminosity 
(e.g. Jansen, Franx, \& Fabricant 2001). 
Because we have no information about metallicity or
dust extinction, and how these might scale with 
luminosity for our sample,
we have not included these corrections. They should therefore be considered lower 
limits on the true SFRs. In the next section we assume an
average internal extinction of 1 magnitude at H$\alpha$ for comparison
with other studies.
We convert our line fluxes to luminosities by assuming all 
galaxies are at the cluster distance ($z=0.84$).

The emission-line fluxes were measured by fitting 
single Gaussian profiles interactively 
using SPLOT in IRAF. We fit several spectra on
five separate occasions to estimate errors on our flux measurements
due to continuum selection.
For the highest line fluxes, the errors were 
$<5$\%, while for the lowest, errors approached 20\%.
Combining this error with the error on the SF prescription,
errors on our SFRs range from $13-50$\%. A representative 
L[OII] error of 10\% results in a SFR error of $\sim 27$\%.
EWs were also measured interactively with SPLOT in IRAF on 
background-subtracted but non-flux calibrated spectra.
EWs were estimated as the mean of 5 separate measurements,
and the quoted errors are the maximum deviation from
the mean measurement.

We can set a $2\sigma$ limit on the SFR for those galaxies where the 
[OII] line was not detected. A typical [OII] emission
line width for the 
star-forming galaxies in our sample is 11\AA.
We set the $2\sigma$ limit as twice 
the standard deviation 
in counts within $\pm 2.5\times11$\AA~of the wavelength of 
the redshifted [OII].
We then relate the mean counts to the mean flux
to arrive at upper limits for SFRs. These are below 1.0~M$_{\odot}~yr^{-1}$
for all galaxies (with two exceptions). For 
comparison, the lowest measured SFR is 0.12~M$_{\odot}$~yr$^{-1}$.
This reflects the widely varying quality of the spectroscopic
sample. The distribution of our SFR limits is shown in Figure~\ref{sflimits}.


The smallest slit width for the spectroscopic observations
is $1\arcsec$.
With the exception of 3 galaxies, the effective radii of all
identified star-forming cluster members are less than $1\arcsec$, 
and the seeing was $< 0.8\arcsec$,
so we do not apply an aperture correction.

\subsection{Integrated Star Formation Rate}
 
Due to the small number of galaxies in our sample and 
redshift completeness effects with color, we choose not to 
compute the fraction of SF galaxies with density or radius
as it is fraught
with uncertainty due to selection effects. To minimize discrepancies 
due to differences in survey characteristics,
\citet{Finnetal04} compared integrated SFRs, which are dominated
by galaxies with the largest star formation rates. A comparison of 
integrated SFRs is more accurate than a fractional comparison.
Here we compare the overall star-forming properties of CL0152-1357
to that of other clusters at low to intermediate redshift.
The spatial limits of the lower redshift
cluster studies determine a limit of $0.5R_{200}$ for comparison.
$R_{200}$ is the radius where the mean cluster density is 200 times the
critical density of the universe, $\rho_{c}$, at that redshift. We write the
expression for $R_{200}$ as 

\begin{eqnarray*}
R_{200}(kpc) = [\frac{M_{cl}}{4/3 \pi 200 \rho_{c}}]^{1/3}
\end{eqnarray*}

We take $M_{cl}$ as the total projected mass within 1~Mpc:
M$_{cl}=4.92\pm0.44\times10^{14}$~M$_{\odot}$ \citep{Jeeetal04}.
For CL0152 at $z=0.84$, $R_{200}=1192$~kpc, or $\sim2.5\arcmin$.


In comparing our results we must include corrections for
completeness and extinction. Also, because this is a 
merging cluster, it is not straightforward
to choose a cluster center. We discuss these issues in
the following sections.

\subsubsection{Incompleteness Correction}

Compared to an imaging survey, a spectroscopic survey
suffers significantly more from incompleteness. We will
correct for this using the redshift completeness as a
function of color and magnitude, but first
we estimate the validity of our incompleteness correction 
by considering the cluster CL0023, which has spectroscopic 
narrow-band imaging data.
70\% of the star forming galaxies and 65\% of the star
formation was missed in the spectroscopic
survey versus the imaging survey covering the same area
\citep{Finnetal04}. Correcting for this incompleteness
should be the same as correcting
for incompleteness in the redshift survey \citep{Letal98}. 
Of the 23 galaxies in the Finn et~al. (2004) sample 
with SFRs $\ge 0.2$~M$_{\odot}$~yr$^{-1}$ that are matched with
spectroscopically confirmed cluster members in the Lubin et~al. 
(1998) sample, 21 are brighter than $R=24.0$. The cumulative 
completeness of the spectroscopic survey is $0.187\pm 0.006$
at $R=24.0$, and $0.240\pm 0.076$ at $R=23.75$. Therefore the
incompleteness is $76-82$\%, in good agreement with the narrow-band 
imaging results. Thus, when we compute the normalized integrated
star formation rate below, we correct for incompleteness
using the redshift completeness function presented in Figure~\ref{selfunc}.

\subsubsection{Extinction Correction}

For the
galaxies where we detect [OII], our star formation rates are 
underestimated due to reddening. From the Nearby Galaxies Survey,
\citet{Jansenetal00} established that there is a luminosity-extinction
relation, in the sense that more luminous galaxies have 
larger $E(B-V)$. However, the scatter in this relation is large,
and it is unclear if intermediate redshift cluster galaxies
follow the same relation as the local star-forming population.
Therefore, to correct for extinction, we follow convention and
assume 1 magnitude of
internal extinction at H$\alpha$, $\lambda 6563$ (e.g. Balogh \& Morris 2000,
Couch et~al. 2001; Balogh et~al. 2002, Finn et~al. 2003). This 
corresponds to 
$\sim 1.8$ magnitudes at $\lambda 3727$ using the Galactic
extinction law of \citet{CCM89}. 
We thus correct our fluxes by a factor of 5.25.

\subsubsection{Integrated SFR for CL0152}

If we restrict ourselves to $0.5R_{200}$, then the choice of 
a cluster center is an important one. The
distribution of cluster galaxies and the X-ray emission are 
elongated. In the spectroscopic survey, more star-forming galaxies 
are detected at the southern 
end of the cluster. In fact, if we choose the center of the northern
clump as $R=0$, then we have no star-forming galaxies within
$0.5R_{200}$. Our integrated SFR changes from 
4.5 M$_{\odot}$~yr$^{-1}$ with 5 galaxies, to 5.3 M$_{\odot}$~yr$^{-1}$ 
with 8 galaxies if
$R=0$ is between the clumps or on the southern clump. If we were
to take $R_{200}$ instead of $0.5R_{200}$, the integrated SFR would not depend significantly
on the choice of cluster center, changing at most
from 19.1 to 19.9 M$_{\odot}$~yr$^{-1}$.

We choose a point between the 
two massive clusters as the cluster center: 01:52:41.55, $-$13:57:56.7.
Our observed, uncorrected integrated star formation rate
is $4.5$~M$_{\odot}$~yr$^{-1}$. Next we correct this
for incompleteness and extinction. Four of these galaxies have 
$r-i < 1.0$, are brighter than $i_{775}=22.8$, and
have a mean SFR of 1.1~M$_{\odot}$~yr$^{-1}$. 
We measure a cumulative completeness of 0.17 for galaxies with $i_{775}\le 23.0$,
and 0.23 for galaxies with $i_{775}\le 22.5$, so we will use a
cumulative completeness of 0.20 for this sample. Then
we should have roughly 20 galaxies with a mean SFR of 0.9~M$_{\odot}$~yr$^{-1}$.
The fifth galaxy has $r-i=1.03$ and SFR=0.2~M$_{\odot}$~yr$^{-1}$. 
The cumulative completeness for red galaxies is 0.53, and we do
not apply a correction. Thus, the completeness 
corrected total SFR is 18.2~M$_{\odot}$~yr$^{-1}$. We then apply an 
extinction correction corresponding to 1.8 magnitudes of extinction 
at [OII]. We thus correct our total SFR by a factor of
5.25, for a total of 96~M$_{\odot}$~yr$^{-1}$. The error on this estimate
is unlikely to be lower than 20\%.

\section{A Comparison of Cluster SFRs}

To investigate what dominates the star-forming
properties of cluster galaxies, evolution or cluster characteristics, 
in Figure~\ref{intsfr}
we plot integrated SFRs and the mass-normalized 
integrated SFRs versus redshift, $L_{X}$, and $T_{X}$. 
Cluster parameters are listed in Table~\ref{tab:clus}.

\citet{Finnetal04} plotted mass-normalized integrated cluster SFRs
versus redshift and velocity dispersion, which was intended as a 
proxy for mass. \citet{Kodamaetal04} added
the cluster CL0024+17 at $z=0.395$ to this plot. Here we
add CL0152 at $z=0.837$, and also plot the integrated
SFR and the mass-normalized cluster SFR against 
two quantities that describe the state of the ICM:
the X-ray luminosity and the X-ray gas temperature. 

We take integrated star formation rates and errors
from \citet{Finnetal04} and \citet{Kodamaetal04}. 
The integrated star formation rates for A1689, AC114, A2390, 
CL0024, CL0023, and CL0152 are $61\pm 19$ \citep{Baloghetal02}, 
$39\pm 15$ \citep{Cetal01}, $96\pm 34$ \citep{BM00}, 
$253\pm 34.$ \citep{Kodamaetal04}, $78\pm 7$  
\citep{Finnetal04}, and $96\pm24$~M$_{\odot}$~yr$^{-1}$, 
respectively. These numbers were derived from H$\alpha$ spectroscopy 
for A1689 and AC114, and
$H\alpha$ narrow-band imaging for A2390, CL0024, and 
CL0023. Here we include an incompleteness
correction of 2.5 for AC114 from the values presented in 
\citet{Finnetal04}, and for A1689 adopt the incompleteness and 
a slit aperture correction of 2.8 from \citet{Kodamaetal04}. 
Where errors on $L_{X}$ are not listed, we assume a 10\% error.

In the right-hand panels, we divide the integrated
SFRs by cluster mass in units of $10^{14}$~M$_{\odot}$.
We refer to these values as mass-normalized cluster
SFRs. For CL0152, CL0024, and AC114 we use the lensing 
masses of $4.9\times 10^{14}$~M$_{\odot}$ \citep{Jeeetal04},
$5.7\times 10^{14}$~M$_{\odot}$ \citep{Kneibetal03}, 
and $7.3\times 10^{14}$~M$_{\odot}$ \citep{Natarajanetal98}.
For consistency with \citet{Kodamaetal04} we use 
$8.1\times 10^{14}$~M$_{\odot}$ for A1689 \citep{KCS02}, but 
we note that recent work by \citet{Broadhurstetal04} and
\citet{Zekseretal04} indicate a value of 
$1.8\times 10^{15}$~M$_{\odot}$. This also illustrates the
typical uncertanties in mass estimates from lensing analyses.
For CL0023 we use the estimate of $2.3\times 10^{14}$~M$_{\odot}$
in \citet{Kodamaetal04}. Where errors on mass are not listed, 
we assume a 10\% error.

We find a possible
inverse correlation of integrated SFR and $T_{X}$, 
but also note that the scatter is large.  Using the non-parametric
Spearman rank correlation test, we find a
70\%, or $1\sigma$, inverse
correlation between
integrated SFR and $T_{X}$. However, this disappears
when CL0024 is excluded. For the mass-normalized cluster SFR 
and $T_{X}$, it is more significant, 94\%. The trend remains
even when CL0024
is excluded. We find no significant correlation between either
the integrated SFR and redshift or the mass-normalized integrated
SFR and redshift.

CL0024 has the highest integrated SFR of the clusters surveyed 
so far. It also has the lowest X-ray temperature and
X-ray luminosity, but its weak-lensing mass (within $M_{200}$)
is similar to CL0152 (within 1~Mpc) and AC114, and it is at a similar redshift to AC114.
Considering only these 3 clusters, it appears that it is the ICM that has the
dominant effect on star formation in cluster galaxies. However,
we caution that until more clusters with low $T_{X}$ and $L_{X}$
are observed, such a conclusion is tentative.
In fact, it may be that $T_{X}$ is a more accurate measure
of cluster mass than the values derived from lensing analyses,
in which case the fundamental correlation is between cluster SFR
and cluster mass, with either no or weak evolution between $z=0.2-0.8$. 
One possible complication is additional ICM 
heating from AGN, which may introduce some scatter in the 
expected T$_{X}^{3/2}\propto$M$_{clus}$ correlation for a virialized 
system. There is some evidence for a significant AGN population in
the $z=0.83$ cluster MS1054 \citep{JBA03}, and CL0152 has two confirmed
AGN associated with the southern clump \citep{Demarcoetal04}.

Let us briefly review what is qualitatively expected from
hierarchical models. \citet{Borganietal02} (see Fig.~6) predict a 
trend of a rising SFR with redshift
until $z\sim2-3$, and a decline thereafter. More massive
halos have larger SFRs at high redshift which rise more slowly,
peak at higher redshift, and decline more sharply.
Around $z\sim2$ the trend of SFR with halo mass has reversed,
with lower mass halos having higher SFRs. There is also
a prediction of a larger spread of integrated cluster 
SFRs with decreasing redshift for clusters that differ in mass,
but an overall trend of decreasing cluster SFR with redshift.
We do not find evidence for this in our data. We note
that if CL0024 is excluded, then there is a tentative correlation 
of mass-normalized cluster SFR with redshift, however, this
arises because in this sample more massive clusters are at 
lower redshift. Also, ``normalizing'' the model halos by mass
exaggerates the differences in cluster SFRs at these redshifts,
and would not give such a correlation. Testing such models 
must involve observations
of many clusters at $z=0-1$ over a range of cluster masses
to look for similarity at intermediate redshifts and strong
diversity at low redshifts.

\section{Star Formation Extinguishment Mechanism}

The segregation of star-forming and passive galaxies in this cluster
is striking. That no star-forming galaxies are found within the 
regions of high ICM density points to a galaxy/ICM interaction 
as the cause. There are two major contenders in the galaxy/ICM
interaction category: ram pressure stripping and starvation. 
The two are distinguished by the timescale over which they are
expected to operate. Ram pressure stripping affects the entire
gas supply of a galaxy, disk and halo, whereas starvation 
only concerns the more loosely bound halo of gas, 
commonly termed the 'reservoir', which if left undisturbed, 
will settle to the disk and participate in
star formation. Ram pressure stripping is expected to occur
on timescales of a few tens of Myr, and starvation $\gtrsim 1$~Gyr.


For a comparison with \citet{Treuetal03},
we estimate, using standard
assumptions, the radius at which
ram pressure stripping becomes important for a typical infalling 
spiral galaxy into the {\it southern cluster}. 
We assume a smooth ICM and a density profile of the form

\begin{eqnarray*}
\rho_{gas}=\rho_{0}\times[1+(r/r_{c})^{2}]^{(-3/2)\beta} \\
\end{eqnarray*}

We use the virial radius $r_{v}=1.4$~Mpc, the gas mass within
the virial radius $4.5\times 10^{13}$~M$_{\odot}$, the core radius,
$r_{c}$=$122^{+28}_{-20}$~kpc, and $\beta=0.66\pm 0.08$ \citep{Metal03}.
to find $\rho_{0}=2\times 10^{14}$M$_{\odot}$~Mpc$^{-3}$. 
Following \citet{FN99}, we write the condition for ram pressure stripping as

\begin{eqnarray*}
\rho_{gas}v^{2}> 3.1\times 10^{19} M_{\odot}~km^{2}~Mpc^{-3}~s^{-2} \\ 
(\frac{v_{rot}}{220 km s^{-1}})^{2}(\frac{r_{h}}{10 kpc})^{-1}(\frac{\Sigma_{HI}}{8\times 10^{20}m_{H}})
\end{eqnarray*}

where $\rho$ is in M$_{\odot}$~Mpc$^{-3}$, $v$ is in km~s$^{-1}$,
and $m_{H}$ is in cm$^{-2}$.

With these parameters, the stripping radius for a Milky Way-type
galaxy moving at a velocity of 1000~km~s$^{-1}$ relative to the ICM
is $\sim 0.3$~Mpc. The closest projected star-forming galaxies 
to the southern clusters are $\sim 0.3-0.4$~Mpc. Although this is a 
crude estimate,
the segregation of 
star-forming and non-star-forming galaxies in CL$0152-1357$ is
consistent with an ICM interaction as the dominant mechanism. 
The crucial 
distinguishing observation is the time for star formation to be 
extinguished, which we cannot make here.

\section{Evolution of Star-forming Galaxies in CL0152}

Each of the red sequence spirals has evidence
for a central stellar concentration above an exponential
disk, which we will refer to as a bulge. This contrasts
with the blue late-type star-forming galaxies, about half of
which show no evidence for a bulge. Their red colors cannot
be due solely to dust, as they appear smooth while
the blue spirals have patchy appearances, a clear indication 
of a larger amount of dust extinction. The fact that these 
red spirals are in the red cluster sequence means that the
bulk of their stellar population was formed at a similar
redshift as the ellipticals that make up the majority of the
RCS in this cluster. Given their red color, 
the presence of a bulge, and
their relatively low star formation rates, it is probable 
that these red sequence spirals will become S0s.

The compact early-type galaxies have rest-frame $M_{B}$
values from $-19$ to $-20.5$. A minimum fading of $-1.2$ 
magnitudes from $z=0.84$ to 0
is expected assuming a single burst population
with an age of 3~Gyr at $z=0.84$. Adding even a 1\%
burst with an age of 10~Myr increases the fading to $-2.2$ magnitudes, 
and a burst of 10\% to $-4.1$ magnitudes. We therefore
expect these galaxies to fade to M$_{B}=-19$ to $-15$. 
This is consistent with an identification
as progenitors of dwarf elliptical
galaxies that are populous in present day clusters 
\citep{CGW01,GGM03,JBB04,RS04}. 


These compact early-type galaxies may also be 
related to ``harassed'' low surface brightness disk galaxies.
The simulations of \citet{Metal99} explore the 
effects of galaxy harassment on high and
low surface brightness (LSB) disk galaxies, and  indicate that the 
outcome is significantly different for the two types of objects. LSB
galaxies lose between $50-90$\% of their stars, and 
remnants are well-fitted by exponential disks with
scale-lengths between $1.5-2.5$~kpc, which is consistent with 
the scale-lengths of the compact early-type galaxies discovered
in CL0152. To match the absolute $B$ magnitudes
of these compact galaxies, infalling field LSB disk galaxies 
would have to have absolute B
magnitudes of $-21$ to $-23$ before the effects of harassment, 
assuming the young and old stellar populations are equally
stripped.

We also note a possible connection to the 
''missing'' faint red galaxy population in the luminosity
functions of intermediate redshift
clusters \citep{Kodamaetal04,Gotoetal05}. Perhaps the 
faint red population is still forming at $z=0.8$, and the
descendants of these star-forming compact galaxies will make
up the missing faint red galaxies.

\section{Conclusions}

We have identified a population of star-forming galaxies
in the cluster CL0152-1357 at $z=0.837$ \citep{Demarcoetal04}, and we investigate
their morphologies, colors, and spatial distribution. We derive star-formation 
rates from [OII] fluxes, and use these to 
compare the integrated SFR within $0.5R_{200}$ 
with other clusters between $z=0.2-0.8$. 
Our conclusions are as follows.

\begin{itemize}

\item The 24 star-forming galaxies with ACS morphological information
range from compact 
early-types to spirals, with only one highly disturbed galaxy. 
The colors of the
star-forming galaxies range from blue to red, and 8 out of 24 have 
$r_{625}-i_{775}$
colors placing them on the red cluster sequence. We find that the
fraction of red cluster sequence galaxies with
[OII] emission is likely to be $>0.15$. Six of these
galaxies have obvious
spiral morphologies, and presuming an absence of major mergers, 
will most likely evolve into red sequence S0s. Thus, $>15$\% 
of red cluster sequence S0s in massive clusters
at z=0 are still forming stars at a low level at $z=0.84$.
However, their overall red colors indicate that they
formed most of their stars during the epoch of massive
cluster ellipticals.

\item We find no massive early-type galaxies with
[OII] emission lines down to a 12~\AA~ observed limit,
corresponding to approximately 7~\AA~ rest-frame EW. 
All galaxies with [OII] are
spiral, irregular, or compact, low-to-intermediate mass
early-type galaxies. These compact early-type galaxies have
rest-frame $B$ magnitudes of $-19$ to $-20.5$. Assuming a 
moderate amount of $1.5-4$ magnitudes of $B$-band fading, this
overlaps with present-day dwarf elliptical magnitudes, and 
we identify them as possible dwarf elliptical progenitors. 
We also note that these may be the ''missing'' faint red galaxy population in 
intermediate redshift clusters \citep{Kodamaetal04,Gotoetal05}.

 \item We find a paucity of star-forming galaxies within $\sim 500$~kpc
of the projected cluster center, and none within the 
main X-ray peaks. Using simple, standard assumptions about the
density profile of the ICM, this radius is consistent with 
the expected effect of ram pressure stripping. 

\item We measure the integrated star formation rate
within $0.5R_{200}$ for CL0152-1357, and find a value
similar to that of the $z=0.8$ cluster CL0023+0423, although
these clusters differ by at a least a factor of 2 in mass.
We find an
inverse correlation of the mass-normalized integrated 
star formation rate and $T_{X}$, and no trend with redshift. 
However,
a larger cluster sample is needed to reach a definitive conclusion,
and additional data on clusters with low X-ray temperatures
will be pivotal.

\end{itemize}

\acknowledgements

ACS was developed under NASA contract NAS 5-32865, and this research 
has been supported by NASA grant NAG5-7697 and 
by an equipment grant from  Sun Microsystems, Inc.  
The {Space Telescope Science
Institute} is operated by AURA Inc., under NASA contract NAS5-26555.
We are grateful to K.~Anderson, J.~McCann, S.~Busching, A.~Framarini, S.~Barkhouser,
and T.~Allen for their invaluable contributions to the ACS project at JHU. 
We thank W.~J. McCann for the use of the FITSCUT routine for our color images.

\begin{deluxetable}{lcccccrcccc}
\tablecolumns{11}
\tablewidth{0pc}
\tablecaption{Properties of the Star-Forming Cluster Members
\label{sftab}}
\tabletypesize{\scriptsize}
\tablehead{
\colhead{ID} & \colhead{ID} & \colhead{$z$}  & \colhead{$r_{625}-i_{775}$} &  \colhead{$i_{775}-z_{850}$} &  \colhead{$i_{775}$} & 
\colhead{$\Delta$ RCS}  & \colhead{Flux[OII]} & \colhead{SFR} & \colhead{EW [OII] (\AA)}\\
\colhead{VLT} & \colhead{ACS} & \colhead{} & \colhead{MAGISO} & \colhead{MAGISO} & \colhead{MAGAUTO} & 
\colhead{mag} & \colhead{erg s$^{-1}$ cm$^{-2}$} & \colhead{M$_{\odot}$ yr$^{-1}$} & \colhead{\AA}}
\startdata
   1290 &  8671   & 0.8416 & $0.715\pm 0.011$ & $0.305\pm 0.010$ & $21.411\pm 0.008$ & $-0.49$ &  1.51E-16  & 3.4 & $-59.9\pm 6.0$\\
   650  &  3927   & 0.8671 & $0.839\pm 0.013$  & $0.333\pm 0.011$ & $21.574\pm 0.009$ & $-0.36$   &  1.32E-16  & 2.9 & $-55.2\pm 14.7$\\
   125  &   \nodata & 0.8376 &  \nodata     &   \nodata     &   \nodata    & \nodata  &  1.23E-16  & 2.7 & $-167\pm 30$\\
   144  &   \nodata & 0.8442 &  \nodata     &   \nodata     &  \nodata      &  \nodata  & 1.05E-16  & 2.3 & $-31.9\pm 6.7$\\
   1006 &  10871  & 0.8485 & $0.440\pm 0.032$ & $0.186\pm 0.038$ & $24.028\pm 0.026$ & $-0.66$ &  1.01E-16  & 2.2 & $-579\pm 422$ & \\
   347  &  1676   & 0.8463 & $0.897\pm 0.014$  & $0.424\pm 0.012$ & $22.725\pm 0.007$ & $-0.25$ &  9.24E-17  & 2.0 & $-64.8\pm 13.1$\\   
   1530 &   \nodata & 0.8367 &  \nodata      &  \nodata      &   \nodata     &  \nodata  & 8.20E-17  & 1.8 & $-185\pm 28$\\
   1146 &  5111   & 0.8641 & $0.762\pm 0.015$ & $0.289\pm 0.013$ & $23.016\pm 0.008$ & $-0.38$ &  7.09E-17  & 1.6 & $-67.7\pm 11.0$\\
   306  &  5481   & 0.8539 & $1.036\pm 0.015$  & $0.634\pm 0.009$ & $22.019\pm 0.008$ & $-0.14$  &  6.55E-17  & 1.4 & $-51.9\pm 35.0$ \\
   3014 &   \nodata & 0.8474 &  \nodata     &   \nodata     &   \nodata     & \nodata & 4.72E-17  & 1.0 & $-29.6\pm 8.9$\\
     295  &  1652   & 0.8370 & $0.803\pm 0.012$  & $0.279\pm 0.011$  & $21.633\pm 0.009$ & $-0.39$ &  4.72E-17  & 1.0 & $-31.2\pm 4.9$\\
     327  &  2016   & 0.8247 & $0.261\pm 0.014$  & $0.120\pm 0.017$ & $22.277\pm 0.011$ & $-0.91$ &  4.24E-17  & 0.9 & $-273\pm 165$ \\
   161  &   \nodata & 0.8447 &  \nodata     &  \nodata      &   \nodata     & \nodata & 4.18E-17  & 0.9 & $-85.6\pm 13.0$ \\
   1131 &  5846   & 0.8237 & $0.403\pm 0.020$ & $0.342\pm 0.021$ & $23.842\pm 0.014$ & $-0.71$ &  3.72E-17  & 0.8 & $-646\pm 522$ & \\
     868  &  8708   & 0.8297 &  $0.573\pm 0.030$  &  $0.263\pm 0.031$ & $23.444\pm 0.025$ & $-0.55$ & 3.09E-17  & 0.7 & $-89.1\pm 40.6$\\
     47   &  2027   & 0.8436 & $0.884\pm 0.019$  & $0.417\pm 0.015$ & $22.134\pm 0.013$ & $-0.29$   &  2.71E-17  & 0.6 & $-17.2\pm 10.3$ \\
   270  &  1562   & 0.8450 & $0.984\pm 0.019$  & $0.505\pm 0.013$ & $21.613\pm 0.014$ & $-0.21$  &  2.21E-17  & 0.5 & $-31.4\pm 4.8$\\
   851  &  5410   & 0.8360 & $1.022\pm 0.012$  & $0.485\pm 0.008$ & $21.592\pm 0.006$ & $-0.17$   &  1.98E-17  & 0.4 & $-12.3\pm 6.2$\\
   898  &  10148  & 0.8300 &  $0.711\pm 0.023$    & $0.296\pm 0.021$ &  $22.774\pm 0.013$ & $-0.44$ & 1.71E-17  & 0.4 & $-205\pm 107$\\
   1532 &  1146   & 0.8413 & $0.740\pm 0.024$ & $0.316\pm 0.022$ & $22.584\pm 0.017$ & $-0.42$ &  1.95E-17  & 0.4 & $-19.5\pm 2.9$\\
   377  &  2597   & 0.8379 & $0.971\pm 0.022$  & $0.465\pm 0.016$ & $22.404\pm 0.014$ & $-0.19$  &  1.67E-17  & 0.4 & $-47.4\pm 7.4$ \\
   1238b&   \nodata & 0.8456 &  \nodata     &   \nodata     &  \nodata      &  \nodata &  1.55E-17  & 0.3 & $-150\pm 30$ \\
   1258 &  7017   & 0.8394 & $1.136\pm 0.010$ & $0.570\pm 0.008$ & $21.250\pm 0.005$ & $-0.07$ &  1.50E-17  & 0.3 & $-13.4\pm 2.2$\\
   184  & \nodata &  0.8397  & \nodata     &   \nodata     &   \nodata     & \nodata & 1.36E-17  & 0.3 & $-73.8\pm 25.0$\\
   204  &  3390   & 0.8386 & $1.247\pm 0.016$  & $0.701\pm 0.008$ & $21.118\pm 0.008$ & $0.04$ &  1.15E-17  & 0.3 & $-20.7\pm 19.5$\\
   248  &  4076   & 0.8472 & $1.111\pm 0.022$  & $0.733\pm 0.013$ & $22.479\pm 0.011$ & $-0.05$ &  1.47E-17  & 0.3 & $-30.9\pm 16.1$\\
   551  &  2235   & 0.8362 & $1.235\pm 0.018$  & $0.604\pm 0.010$ & $21.705\pm 0.009$ & $0.05$   &  1.45E-17  & 0.3 & $-10.6\pm 5.0$\\
   18   &  717    & 0.8248 & $1.263\pm 0.031$  & $0.597\pm 0.017$ & $22.435\pm 0.016$  & $0.10$ &  8.63E-18  & 0.2 & $-94.6\pm 58.1$\\
   234  &  1564   & 0.8474 & $0.784\pm 0.024$  & $0.329\pm 0.021$ & $22.661\pm 0.017$ & $-0.37$ &  7.07E-18  & 0.2 & $-32.6^{+37.6}_{-16.3}$\\
   267  &  1575   & 0.8443 & $1.057\pm 0.039$  & $0.704\pm 0.023$ & $22.876\pm 0.024$ & $-0.09$  &  8.38E-18  & 0.2 & $-139\pm 126$    \\
   394  &  1737   & 0.8329 & $1.029\pm 0.023$  & $0.415\pm 0.016$ & $22.789\pm 0.014$ & $-0.12$  &  6.88E-18  & 0.2 & $-20.7\pm 9.7$\\
   3013 &   \nodata & 0.8224 &  \nodata     &   \nodata     &   \nodata     & \nodata & 5.50E-18  & 0.1 & $-27.0\pm 11.0$\\
   26b\tablenotemark{1}  & \nodata   & 0.8372 & \nodata     & \nodata        &   \nodata    & \nodata   & \nodata  &  \nodata  & \nodata \\
\enddata
\tablecomments{Notes: Colors are computed with MAGISO; total $i_{775}$ is MAGAUTO. $\Delta$~RCS is deviation in magnitudes of $r-i$
from the red cluster sequence. Errors on Flux[OII] ranges from 5-20\% from the
highest to the lowest luminosities. Combining this with the error in the L[OII]-SFR relation \citep{KGJ04}, errors
on SFRs range from $\sim 13-50$\%. EWs are the mean of 5 interactive measurements in SPLOT. EW errors are
taken as the largest deviation from the mean. EWs are {\it observed} EWs; to transform to rest-frame, divide by (1+z).}
\tablenotetext{1}{The blue wavelength coverage was not sufficient to detect [OII] $\lambda$~3727; only [OIII] $\lambda$~4959,5007 was detected.}
\end{deluxetable}

\clearpage

\begin{table}
\begin{center}
  \caption{Structural Parameters of Compact Early-type Star-forming Galaxies \label{tab:compacts}}
  \begin{tabular}{ccccr}\hline
ACS ID & $r_{e}$, pixels  & $r_{e}$, kpc & $n$ & T type\\\hline  
4076  & 7.1  & 2.7 & 4.7  & $-4$ \\
5111  & 4.3  & 1.7 & 1.4  & $-1$ \\
10871 & 3.8  & 1.4 & 0.8  & $0$  \\
1676  & 3.1  & 1.2 & 5.4  & $-5$ \\
5846  & 2.8  & 1.1 & 1.8  & $-1$ \\ \hline
\end{tabular}
\end{center}
\end{table}

\clearpage

\begin{table}
\begin{center}
  \caption{\label{tab:clus}}
  \begin{tabular}{cccccc}\hline
Cluster & Redshift & $10^{14}$ M$_{\odot}$ & $L_{X}$ 10$^{44}$ erg s$^{-1}$ & $T_{X}$ (keV) & Ref\\\hline  
A1689   &  0.183     &  8.1    & $55.73\pm8.92$     &  $9.02^{+0.4}_{-0.3}$ &  1 \\
A2390   & 0.228      &  $13.6\pm0.7$    & $63.49\pm14.87$    &  $11.5\pm1.5$        & 2 \\
AC114   & 0.32       &  $7.3^{+4.4}_{-1.9}$     & $38.10$            &  $9.76^{+1.04}_{-0.85}$ & 3 \\
CL0024  & 0.395      &  $5.7\pm1.1$    & $3.4$              &  $4.47^{+0.83}_{-0.54}$ & 4\\
CL0152  & 0.837      & $4.9\pm0.4$ &  $16.0\pm2.0$ &  $7.7\pm2.0$  & 5\\      
CL0023  & 0.845      &  $2.3$      &   \nodata            &    \nodata           & 6\\
\end{tabular}
\tablerefs{(1) \citet{KCS02,WXF99}, (2) \citet{AEF01}, (3) \citet{WXF99,Natarajanetal98} (4)\citet{Kneibetal03,Otaetal04}, (5) \citet{Demarcoetal04,Jeeetal04,Metal03}, Balestra et~al., in preparation, (6) \citet{Finnetal04,Kodamaetal04} }
\end{center}
\end{table}

\clearpage

\begin{figure}
\begin{center}
\includegraphics[width=16cm]{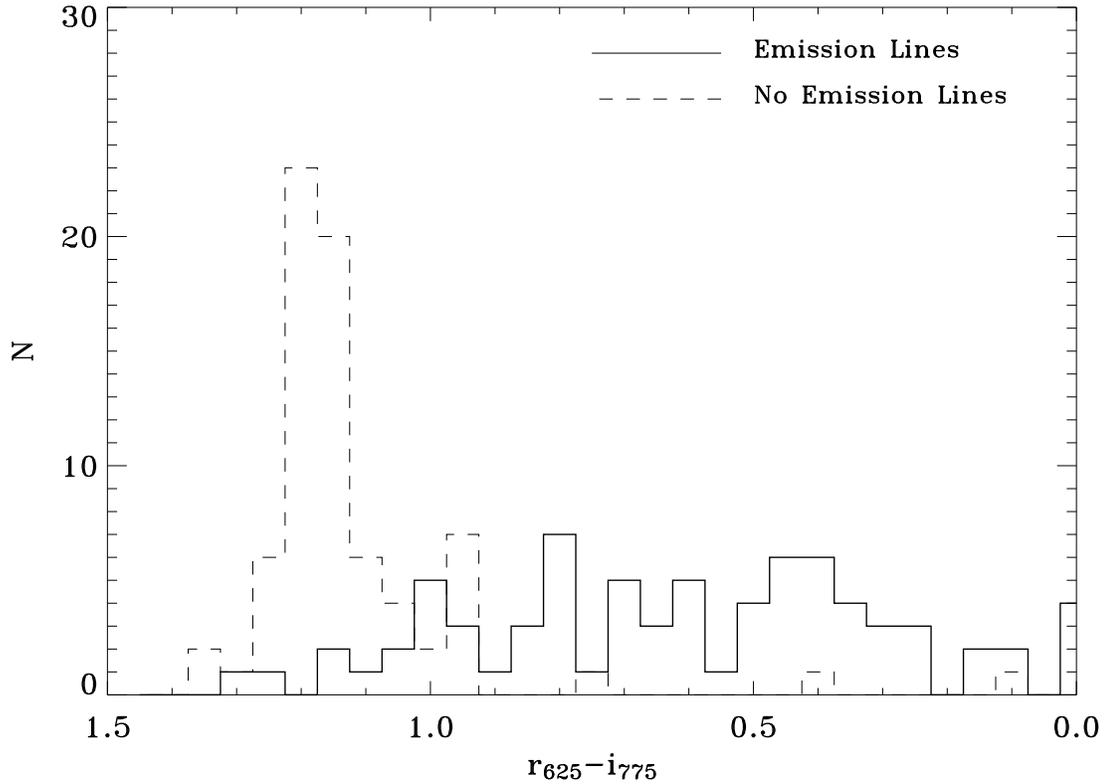}
\caption{Histogram of $r_{625}-i_{775}$ values for cluster members with and without 
detected star formation. Some of the red galaxies have star formation, but the
color distribution of passive galaxies is more peaked than for the star-forming galaxies.
\label{colors}}
\end{center}
\end{figure}

\begin{figure}
\begin{center}
\includegraphics[width=16cm]{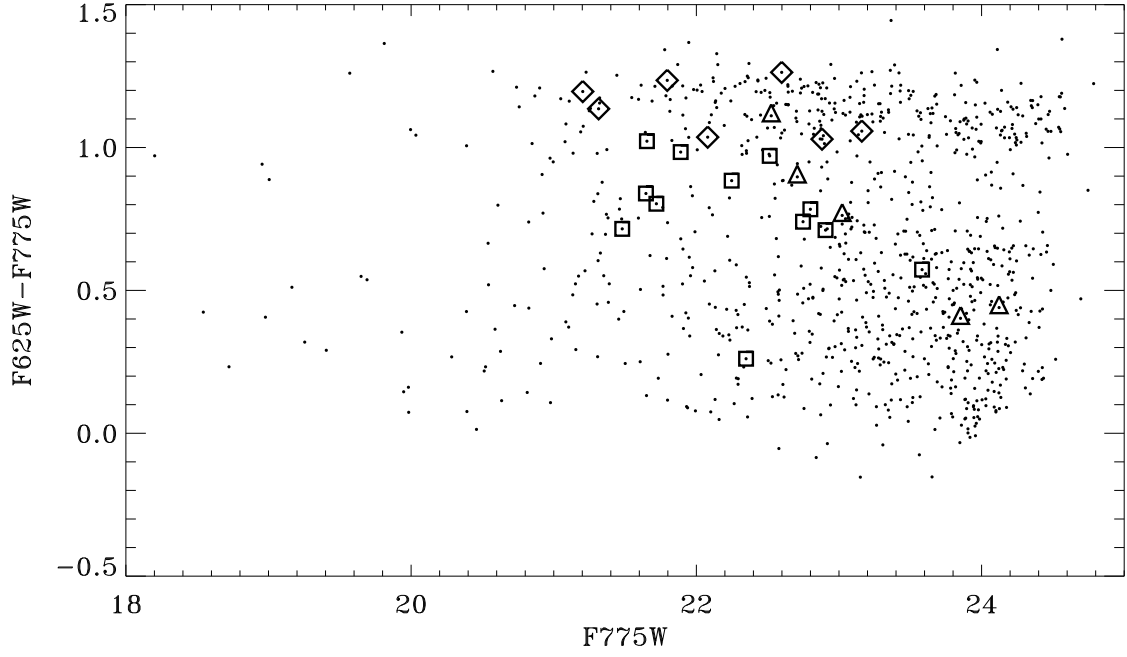}
\caption{Color magnitude diagram for galaxies in the Cl0152 field. Cluster 
member with star formation are marked with triangles, squares, and diamonds for
compact early-type galaxies, blue spirals/irregulars, and RCS spirals, 
respectively.
\label{cmd}}
\end{center}
\end{figure}

\begin{figure*}
\begin{center}
  \includegraphics[width=16cm]{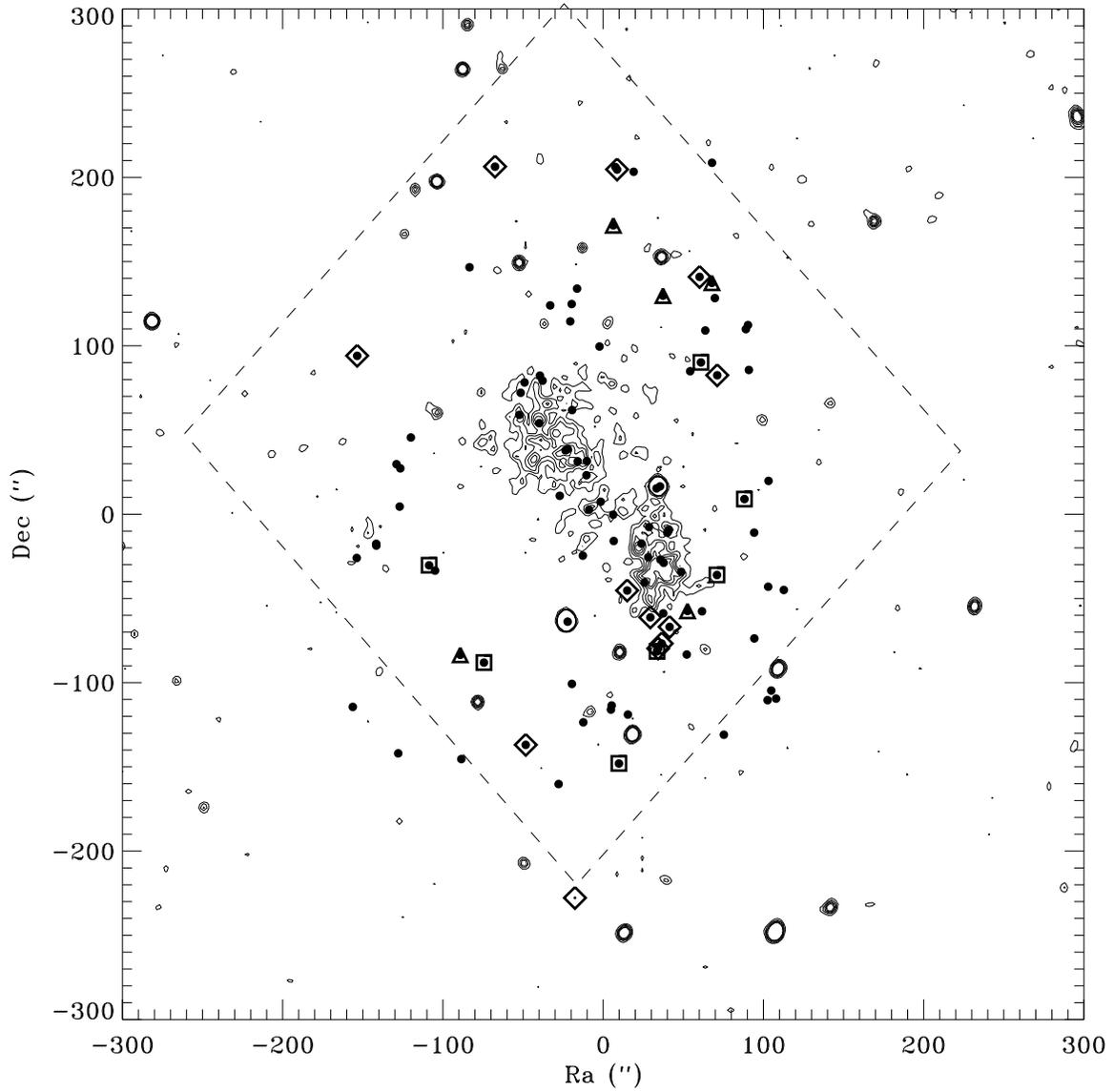}
  \caption{An X-ray contour map from {\it Chandra} with the positions
of the spectroscopically confirmed cluster members plotted as 
filled circles. X-ray contours are drawn at 0.3, 0.5, 0.7, 0.9, and 1.1
counts s$^{-1}$. Star-forming cluster galaxies in the ACS sample 
are indicated as triangles, squares, and diamonds for compacts,
spirals/irregulars, and RCS spirals, respectively.
There is a clear spatial segregation between the cluster members
with star formation and those without. The center of the diagram
is 01:52:42.26, -13:57:57.5 (J2000).
    \label{pos}}
\end{center}
\end{figure*}

\begin{figure}
\begin{center}
\includegraphics[width=8cm]{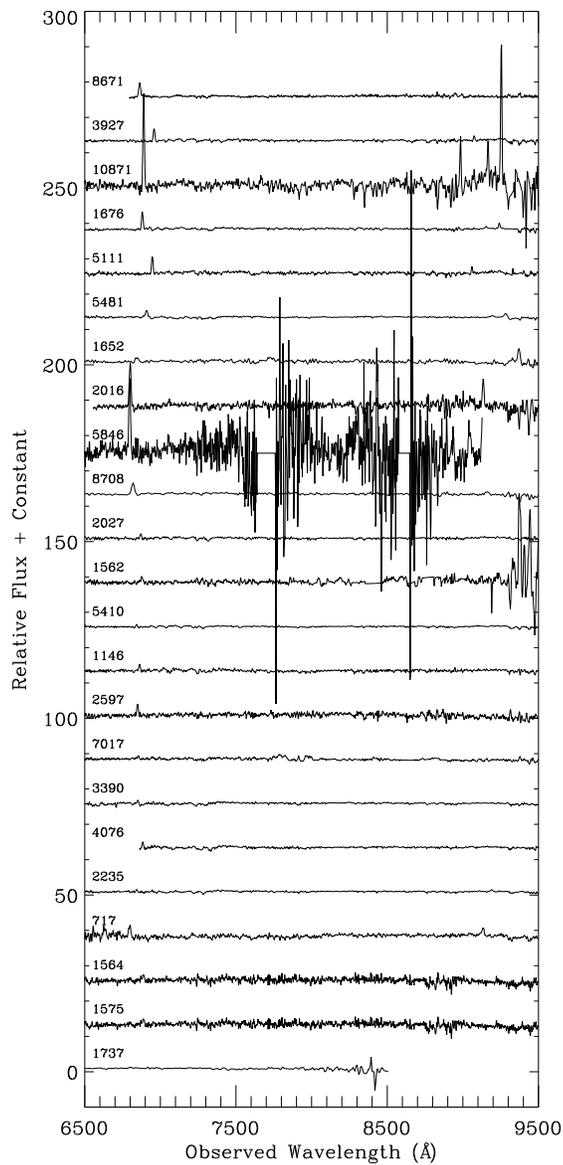}
\caption{Examples of spectra. ACS ID is listed above each spectrum.
\label{spec_examples}}
\end{center}
\end{figure}

\begin{figure}
\begin{center}
\includegraphics[width=16cm]{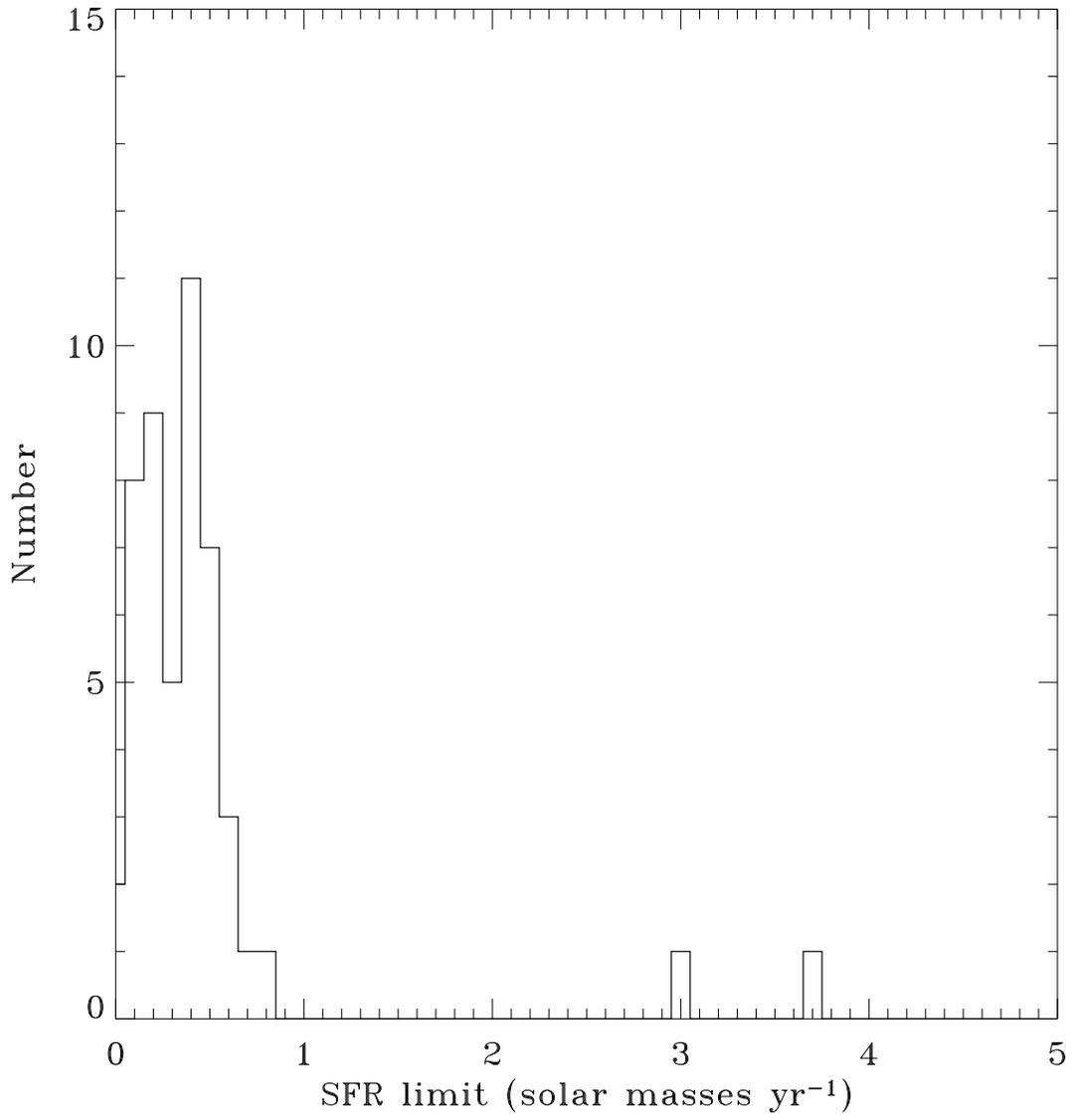}
\caption{Histogram of the $2\sigma$ limits on the star formation
rates for galaxies without detected [OII] emission. All are
below 1~M$_{\odot}$ yr$^{-1}$, with the exception of two galaxies.
\label{sflimits}}
\end{center}
\end{figure}

\begin{figure*}
\begin{center}
\includegraphics[width=16cm]{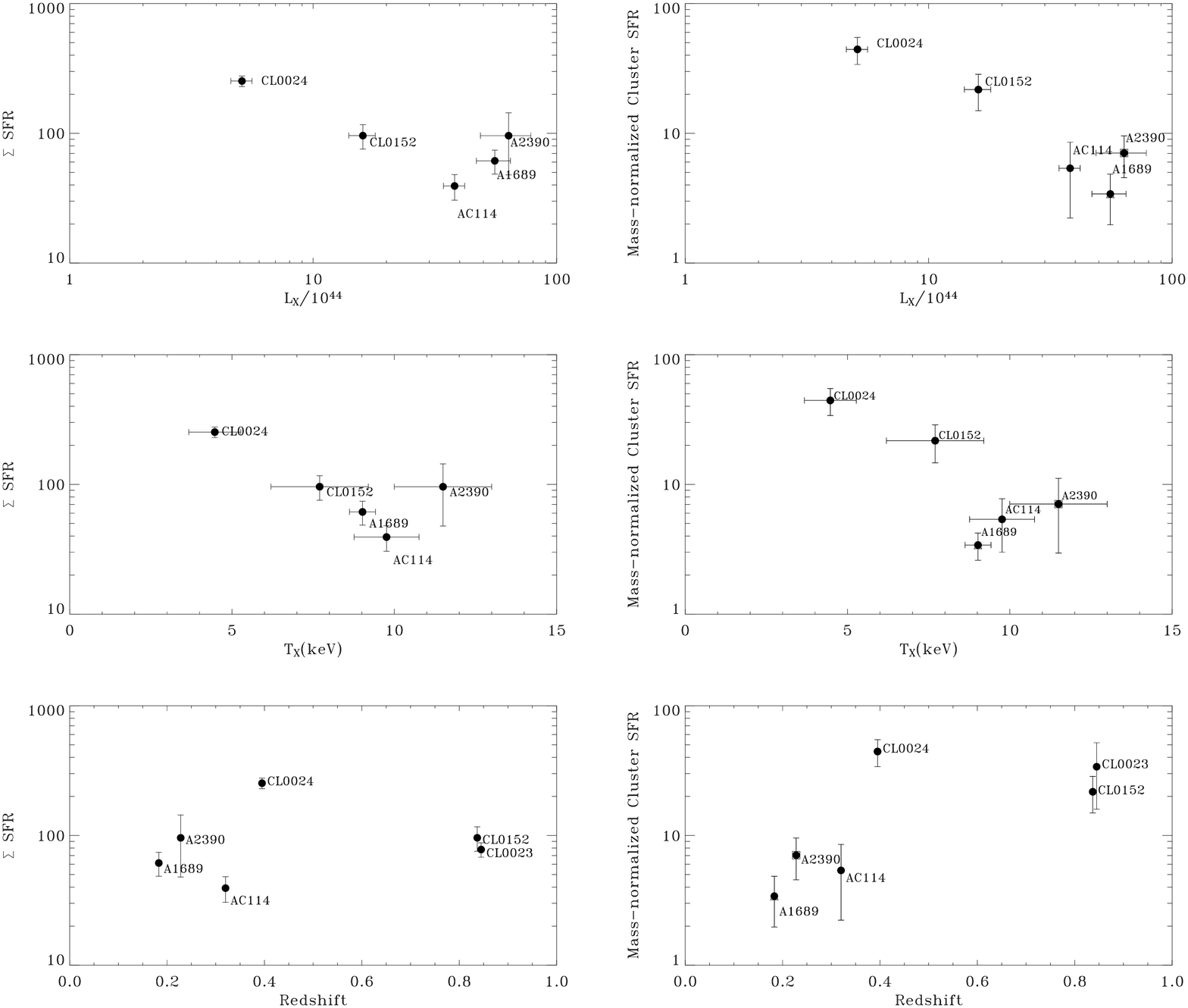}
\caption{Integrated cluster SFRs and mass-normalized integrated cluster SFRs
plotted against $L_{X}$, $T_{X}$, and redshift. See text for details.
\label{intsfr}}
\end{center}
\end{figure*}

\end{document}